\documentclass[twocolumn,10pt]{IEEEtran}
\usepackage{algorithm,algorithmic,amsbsy,amsmath,amssymb,epsfig,bbm,mathrsfs,fancyhdr,cases,comment,color}
\usepackage{rotating,footnote,epstopdf}
\usepackage{multirow}
\usepackage[doublespacing]{setspace}
\usepackage{epstopdf}
\usepackage{graphicx}
\usepackage{caption}
\usepackage{subcaption} 
\usepackage[normalem]{ulem}

\renewcommand{\baselinestretch}{1.0}

\newcommand{\beq}{\begin{equation}}
\newcommand{\eeq}{\end{equation}}
\newcommand{\beqn}{\begin{eqnarray}}
\newcommand{\eeqn}{\end{eqnarray}}

\begin{document}

\title{Ambient RF Energy Harvesting in Ultra-Dense Small Cell Networks: Performance and Trade-offs}
\author{Amin Ghazanfari, Hina Tabassum, and Ekram Hossain\\\thanks{The authors are with the Department of Electrical and Computer Engineering at the University of Manitoba, Canada (emails: ghazanfa@myumanitoba.ca, \{hina.tabassum, ekram.hossain\}@umanitoba.ca). This work was supported by a CRD grant from the Natural Sciences and Engineering Research Council of Canada (NSERC).}}

\maketitle

\begin{abstract}
In order to minimize electric grid power consumption,
energy harvesting from ambient RF sources is considered
as a promising technique for wireless charging
of  low-power devices.
To illustrate  the  design considerations of RF-based ambient energy harvesting networks, this article first points out the primary challenges of implementing and operating such networks, including non-deterministic energy arrival patterns, energy harvesting mode selection, energy-aware cooperation among base stations (BSs), etc. 
A brief overview of the recent advancements and a summary of their shortcomings are then provided to highlight existing research gaps and possible future research directions. To this end, we 
investigate the feasibility  of implementing RF-based ambient energy harvesting in ultra-dense small cell networks (SCNs) and examine the related trade-offs in terms of the energy efficiency and signal-to-interference-plus-noise ratio (SINR) outage probability of a typical user in the downlink. Numerical results demonstrate the significance of deploying 
a mixture of on-grid small base stations (SBSs)~(powered by electric grid) and off-grid SBSs~(powered by energy harvesting) and optimizing their corresponding proportions as a function of the intensity of active SBSs in the network.
\end{abstract}
{\keywords Ambient RF energy harvesting, 5G networks, ultra-dense small cell networks, co-channel interference, outage performance, energy efficiency.}

\begin{figure*}
\centering
\includegraphics[width=0.5\linewidth]{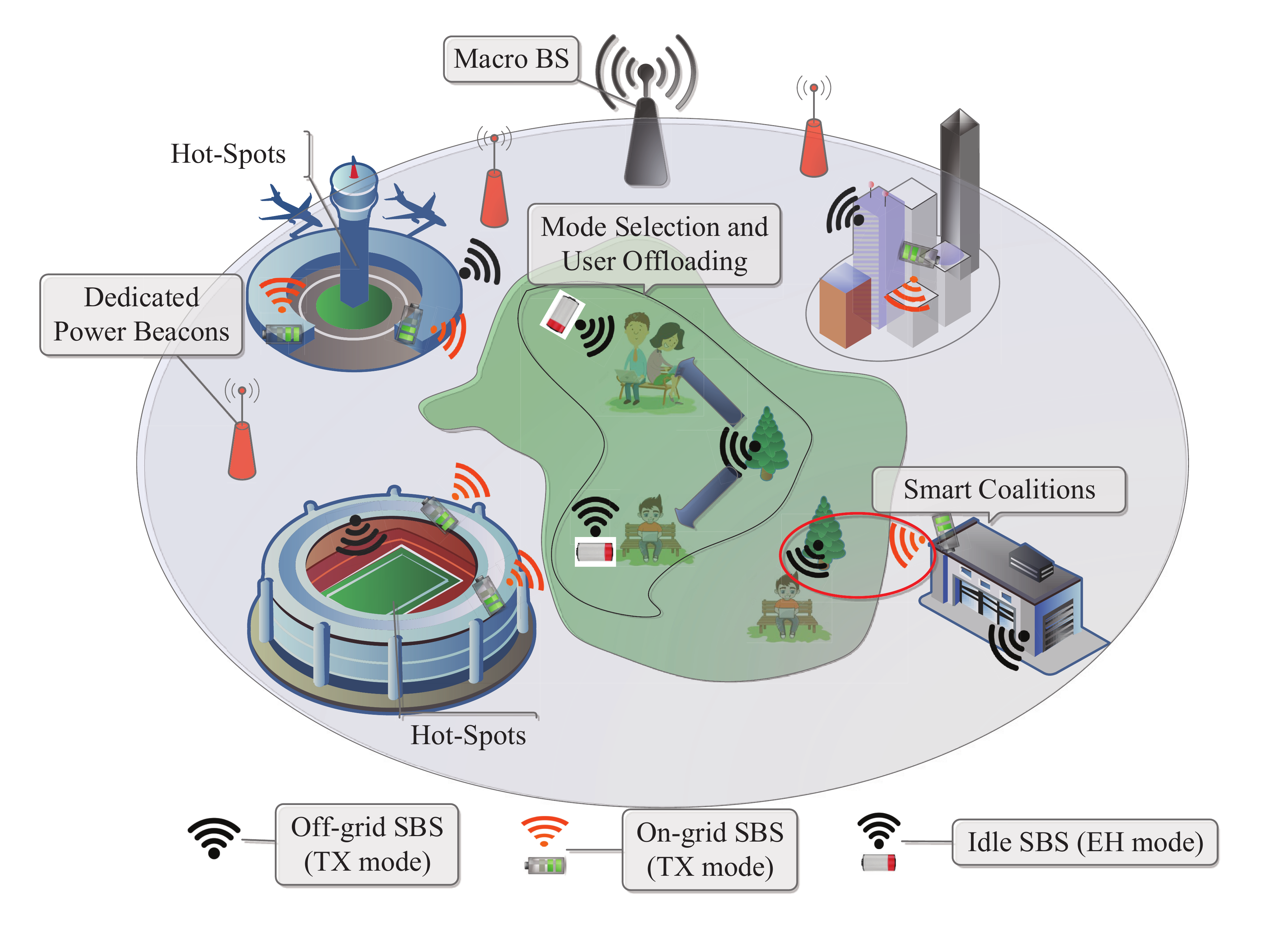}
\caption{An energy harvesting SCN with both dedicated (e.g., macro BSs, dedicated power beacons) and  ambient energy harvesting capabilities (e.g., from co-channel RF transmissions). The SCN comprises of a mixture of both on-grid and off-grid SBSs. Various modes of SBSs (i.e., idle mode, off-grid mode, on-grid mode) coexist that enable smart coalitions and energy-aware user offloading mechanisms.}
\label{fig:sys}
\end{figure*}
\section*{Introduction}
\label{First:intro}
Ultra-dense small cell networks~(SCNs)  are envisioned as a key enabling feature of fifth generation (5G) wireless cellular networks that can potentially meet the high  capacity requirements in outdoor/indoor  environments~\cite{5G}. In such a network, the spatial density of small cells can be 10-100 times of that of traditional macro cells. The successful implementation of such ultra-dense SCNs is challenged by several issues. For instance, the increase in co-channel interference (CCI) due to densification of small cells can significantly degrade the achievable network capacity. Moreover, the subsequent increased energy consumption of the system is not attractive from both the environmental and economical perspectives. Finally, providing grid power to all small cell base stations (SBSs) may not always be feasible due to their possible outdoor/remote/hard-to-reach locations. 

Thanks to the recent advancements in wireless energy harvesting (EH) techniques, it has become feasible to power small  devices wirelessly. Wireless energy harvesting thus enables dense deployment of the SBSs irrespective of the availability of grid power connections. 
In general, wireless energy harvesting can be classified into the following two categories:
\begin{itemize}
\item {\em Ambient Energy Harvesting:}  It refers to the energy harvested from renewable energy sources (such as thermal, solar,  wind) as well as the energy harvested from the radio signals of different frequencies in the environment that can be sensed by EH receivers (e.g., CCI, TV or radio broadcast). 
\item {\em Dedicated Energy Harvesting:} It enables  on-purpose transmission of
energy from dedicated energy sources to EH devices.
\end{itemize}
Since dedicated energy harvesting leverages on the deployment of dedicated energy sources, an additional resource/power consumption is unavoidable~\cite{Ref_ID_33}. Therefore,  ambient energy harvesting is a potential solution to reduce grid power consumption by ultra-dense SCNs.  Unfortunately, due to the  dependence of energy harvested from renewable energy sources on temporal/geographical/environmental circumstances,  consistent performance at the wireless SBSs may not be guaranteed. Also, harvesting energy from renewable energy sources may require an extra hardware set-up of solar panels and/or wind turbines. Thus, in order to minimize grid power consumption of ultra-dense SCNs, the significance of investigating other kinds of ambient energy sources becomes evident.  

RF-based ambient energy harvesting is a potential technique that can support both the higher energy levels and lower grid power consumption in ultra-dense SCNs where significant co-channel transmissions/interferences are likely to exist.  While higher CCI can augment the amount of harvested energy,  it can significantly deteriorate the achievable signal-to-interference-plus-noise ratio (SINR) levels at a wireless device.  It is thus crucial to investigate this trade-off  for design and deployment of RF-based ambient EH networks. 
In this context, our contributions can be summarized as follows:
\begin{itemize}
\item We point out some of the existing and anticipated challenges associated with the  implementation and operation of ambient RF energy harvesting in SCNs.
\item Next, we provide a brief overview of the existing literature in the context of the discussed challenges. The provided review highlights the research gaps and points out future research directions.
\item We 
investigate the feasibility of RF-based ambient energy harvesting in ultra-dense SCNs and analyze the trade-off between  {\em energy efficiency} and {\em SINR outage} of a typical user in the downlink. Energy efficiency is defined as the ratio of achievable downlink data rate of a typical user and the corresponding grid power consumption. The considered network comprises  of a mixture of on-grid\footnote{An on-grid SBS refers to an SBS that utilizes fixed electric grid power for downlink transmission, whereas an off-grid SBS refers to an SBS that utilizes harvested energy for downlink transmission.} and off-grid SBSs~(see Fig.~1 for a graphical illustration). To accurately model the performance of ultra-dense SCNs, we utilize a dual-slope path-loss model and show its significance and accuracy over conventional single-slope path-loss model.
\end{itemize}

Numerical results demonstrate the impact of applying dual-slope path-loss model, significance of deploying 
a mixture of on-grid SBSs~(powered by electric grid) and off-grid SBSs~(powered by energy harvesting), and optimizing their corresponding proportions as a function of the intensity of active SBSs in the network.

\section*{Challenges Associated to RF-Based Ambient Energy Harvesting Networks}
\label{second:challenges}
The idea of  harvesting energy for wireless communications devices from renewable energy sources has been popular since last few years. However, the amount of harvested energy remains limited by the geographical, seasonal, and environmental situations. Recently, there has been a shift toward considering RF-based ambient energy harvesting in order to accumulate energy reliably in a variety of environmental settings. Nevertheless, RF-based ambient energy harvesting has its own limitations that need to be tackled before implementing such a system in practice. In this context, some of the main challenges are discussed below.

\subsection*{Implementation  of Energy Harvesting Receivers}
The hardware implementation of EH receiver circuits is a fundamental challenge for the deployment of dedicated as well as ambient EH networks~\cite{Ref_ID_18}. 
The desired receiver sensitivity for wireless energy harvesting~(e.g., -10~dBm) is quite different from  the desired receiver sensitivity (e.g., -60~dBm) for data transmission. The high  sensitivities of EH receivers can result in  significant  fluctuations in energy transfer due to varying channel fading environments as well as the  relative mobility of energy sources and EH receivers. Thus, efficient EH receiver circuits are required that can operate reliably at reduced sensitivity levels. In addition, voltage multiplier (i.e., a circuit which converts RF signals into DC)  is a main component of RF energy harvester that is implemented by using diode technology. As the level of received RF power is very small, existing circuit technologies should be further advanced to enhance RF-to-DC conversion efficiency of voltage multipliers. 

\subsection*{Uncertain Energy Arrival Rate}\label{chal1}
The degree of uncertainty in RF-based ambient energy harvesting at SBSs is quite low compared to energy harvesting from renewable energy sources. The reason is that the locations and traffic patterns of the SBSs deployed in a given area remain relatively fixed over time. Nonetheless, a reliable energy transfer
may not always be guaranteed due to adaptive transmission policies of the SBSs. To analyze the performance of such systems, it is essential to precisely model the energy arrival rate at EH devices.
Also, in such a system, a 100\% deployment of the off-grid SBSs may not be feasible due to insufficient energy harvested from  the less-dense fixed power TV/radio broadcast towers, macro BSs, and the
reduced power transmissions of the other off-grid SBSs. 
Thus, to maintain a balance between grid energy consumption and the achievable communication performance, the ultra-dense SCNs should comprise of a mixture of on-grid and off-grid SBSs. The proportion of on-grid and off-grid SBSs should be selected to maximize the system throughput/rate while minimizing the grid power consumption. 

\subsection*{Modeling and Optimizing Co-Channel Transmissions}\label{chal2}
Co-channel RF transmission is a fundamental performance limiting factor of the conventional cellular networks and its impact will be more significant in ultra-dense SCNs. However, strong co-channel transmissions can be useful to harvest energy in EH-enabled SCNs. 
For example, in the downlink, it is crucial to manage the trade-off between the amount of harvested energy at an off-grid SBS and the amount of incurred  interference at a given user. This can be done by optimizing the intensity of active SBSs or the proportion of on-grid and off-grid SBSs such that network throughput, grid power consumption, and in turn energy efficiency can be improved. Note that the amount of harvested energy and the CCI are crucial  for performance analysis of ambient RF-based EH SCNs. Thus,  analytical models need to be developed to characterize the co-channel transmissions of SBSs in various network deployment scenarios.

\subsection*{Energy Cooperation/Coalition Among SBSs}\label{chal3}
Traditionally, cooperation among multiple BSs is carried out to enhance the  diversity of transmitted signals and to perform interference mitigation.
However, in EH-enabled ultra-dense SCNs, cooperation may also be required to overcome the energy imbalance among various on-grid/off-grid SBSs. 
This can be done by allowing nearby SBSs to cooperate by sharing  their energy states/requirements/transmission policies with each other and
by enabling cognitive decision-making capability at each SBS about the consumption or utilization of its harvested energy. For instance, by knowing the transmission pattern of nearby SBSs, an off-grid SBS can estimate the amount of energy it can harvest. Then, with limited information exchange, the off-grid SBS can likely encourage its nearby SBSs to continue their transmission if the overall utility of their coalition can be increased.

\subsection*{Enabling On-grid/Off-grid/Idle Mode Selection}\label{chal4}
Due to uncertain energy arrival rate, each SBS in RF-based ambient EH networks need to be enabled with grid connections unless there is a deployment/implementation constraint. This enables an SBS to avoid severe transmission outages due to insufficient harvested energy by operating adaptively in on-grid, off-grid, and idle modes. As has been discussed before, there could be a central entity that can optimize the proportion of on-grid, off-grid, and idle SBSs in the network and forward the decision to all SBSs. The SBSs then switch their mode accordingly. However, this method may impose significant signaling/information overhead and cannot be scalable for ultra-dense deployments. On the other hand, fully distributed mode selection allows an SBS to decide its mode individually in order to enhance its own utility; however, it may degrade the overall system utility. Therefore, a semi-distributed mode selection would be more attractive. 

\subsection{Energy-Aware User Offloading}
Energy cooperation among SBSs allows SBSs to operate in coalitions such that their overall utilities can be maximized. In this context, energy-aware user offloading can be of significant importance. For example, an SBS may switch to grid power due to its reduced level of harvested energy; however, the SBS could possibly coordinate with the neighboring SBSs to handle its traffic load and move into the idle mode instead. This can potentially improve the overall utility of  the coalition by reducing the grid-power consumption as well as interference to other SBSs.

\section*{An Overview of Existing Literature}
\label{third:overview}

In this section, we provide a brief overview of the recent literature in the context of challenges mentioned in Section~\ref{second:challenges}. More precisely, we focus on the references that are related to ambient RF energy harvesting in different communication networks/scenarios such as cognitive radio networks, wireless sensor networks, device-to-device (D2D) communication, and small cell networks. To highlight the existing research gaps, we comparatively analyze the literature considering their problems/assumptions, energy arrival models, their solution methods, the benefits and limitations of ambient energy  transfer in such communication networks. A qualitative summary of this comparison is provided in Table \ref{tb:Table1}.

\subsection*{Modeling Energy Arrival Rate}
For guaranteed quality-of-service (QoS) in ambient RF-based EH-SCNs, precise modeling of RF energy arrivals (co-channel transmissions)  is crucial. This depends on the accurate modeling of the transmit power, locations, and density of the interfering BSs in various deployment scenarios. In this context, a variety of energy arrival models have been recently considered to analyze the performance of RF-based EH networks.

\subsubsection*{Arbitrary energy arrival}

In \cite{Ref_ID_20}, ambient EH (modeled using a sequence of arbitrary i.i.d. random variables) has been exploited for mobile ad hoc networks. Given the energy arrival statistics, network  throughput is maximized by optimizing the transmission powers of transmitters. For numerical results,
the energy arrival rate is modeled by chi-squared distribution.  
However, the impact of energy harvested from the other transmitters in the system is not considered. \cite{Ref_ID_27} considers maximizing the average throughput of the secondary network (SN) by optimizing spectrum sensing time (i.e., the time to detect unused primary spectrum) and sensing threshold of the secondary transmitters (STs).
The energy arrival at the STs is modeled by an i.i.d. random process.

\subsubsection*{Poisson point process~(PPP)-based energy arrival}
In~\cite{Ref_ID_25}, the cellular users harvest energy from the downlink transmission of $K$-tier BSs and use the harvested energy for their uplink data transmission. Since the spatial distribution of BSs is modeled by independent PPPs, energy arrival rate depends on the parameters of the PPPs. By using  tools from stochastic geometry and queuing theory, closed-form expressions are derived for transmission probability (i.e., the probability that a user's battery has harvested sufficient energy for transmission), coverage probability (i.e., the probability that the SINR at the receiving BS  is higher than the required threshold) and success probability (i.e., the probability that both transmission and coverage probabilities are satisfied) of a typical user.  
Ambient energy harvesting is also investigated in~\cite{Ref_ID_30}, where STs harvest energy from primary transmitters (PTs). Since PTs and STs are modeled by homogeneous PPPs, the energy arrival rate depends on the PPP for the PTs. Applying tools from Markov chain theory, transmission probability of STs and the SINR outage probability of both the primary and secondary networks are derived.

\subsubsection{Ginibre-determinantal point process (DPP)-based energy arrival} 
RF-based ambient harvesting is investigated in the point-to-point uplink~\cite{Ref_ID_9} as well as downlink transmissions~\cite{Ref_ID_28} of wireless sensor networks. The spatial distribution of the cellular transmitters and in turn the RF energy arrival depends on Ginibre-determinantal Point Process (DPP). Closed-form expressions for the mean and variance of energy arrival rate are derived. In addition, upper bounds are derived on both the power outage probability (i.e., the probability that the harvested energy is not sufficient) and transmission outage probabilities (i.e., the probability that the transmission rate of a sensor node is below the desired threshold).

\subsection*{Network Operation and Energy Management Issues}

In~\cite{Ref_ID_6}, different tiers of BSs are modeled as homogeneous PPPs with varying energy harvesting rate, storage capacity, etc. The energy arrival process is modeled by a PPP.  Each BS decides individually to operate in either active or inactive mode based on its energy arrival rate and energy level of its battery. When a BS decides to be inactive, it increases the load on the neighboring BSs and consequently decreases their performance. The availability of a BS  in a network tier (i.e., the time duration during which a BS is active) is derived and then the availability region is jointly maximized for a general set of BSs in a network tier. The effect of availability region on the coverage probability and  downlink rate of a typical user is also investigated. 
In~\cite{Ref_ID_42}, CCI (which is completely known at the receiver and modeled arbitrarily) is considered to power single antenna receiver in a point-to-point wireless link. The receiver opportunistically harvests energy or receives information depending on the channel and interference conditions. The authors derive the optimal switching mode for the receiver considering delay-limited and delay-tolerant information transmission cases.

In \cite{ACAS2014}, the authors propose two approaches to harvest energy from cyclic prefix (CP) in OFDM receivers to provide energy requirement for the signal processing of the receiver. The first approach is an ambient RF based scheme. It is implemented by modifying the receiver architecture such that the receiver is able to harvest energy from CP. The feasibility of this approach  is shown in terms of power consumption at the receiver. The second approach is a dedicated energy harvesting scheme where the transmitter controls the amount of energy in the CP to regulate the harvested energy at the receiver. This approach is observed to be feasible and it provides a self-sustainable receiver. 
In \cite{OPOW2015}, the authors investigate energy harvesting in a time-varying fading environment for point-to-point transmission between a sensor and its sink considering a single-slope path-loss model. It is assumed that the sink has grid power supply and the sensor harvests energy from ambient RF resources.  A closed-form expression is derived for the distribution of harvested energy. For delay-insensitive traffic, the average packet delay is analyzed, whereas for delay-sensitive traffic, packet loss probability is analyzed.

The authors in \cite{EHWC2015}  propose an energy cooperation  scheme for a point-to-point network when both transmitter and receiver have non-idle circuit, i.e., hardware power consumption is not negligible. Both of them harvest energy from external sources to support their communication and the circuit power consumption. In addition, they are able to exchange some of their harvested energy to enhance the communication performance. This paper analyzes the optimal throughput and outage probability for additive white Gaussian noise (AWGN) and Rayleigh block fading channels, respectively. In the former case, the energy arrival is deterministic, whereas in the latter case, it is modeled by a Gamma distribution. Simulation results show that energy cooperation between the transmitter and receiver is crucial for enhanced communication performance.

\begin{savenotes}
\begin{table*}[t]
\scriptsize
\caption{Overview of existing works on RF ambient energy harvesting}
\label{tb:Table1}
\begin{center}
\begin{tabular}
{|p{3.3cm}|p{2.8cm}|p{2.5cm}|p{3.2cm}|p{4cm}|} 
\hline 
& \textbf{Energy arrival model}
& \textbf{Solution techniques} 
& \textbf{Limitations}
& \textbf{Objectives \& benefits}
\\
\hline
MANETs  \cite{Ref_ID_20}
& Arbitrary random process\newline(Chi-squared distribution)
& Homogeneous PPP \newline
Random-walk theory
& EH from CCI is ignored.\newline
Single-slope path-loss 
& Challenge \ref{chal1}
\newline
Network throughput is maximized 
\\
\hline
Cognitive radio \cite{Ref_ID_27}\newline
STs harvest energy from PTs
&Arbitrary random process (i.i.d.)
&Optimization
&CCI of ST is ignored\newline
 Simple i.i.d. for energy arrival \newline
Channel model is not considered
&Challenge \ref{chal1}\newline
Maximize the average throughput of SN
\\
\hline
Cognitive radio  \cite{Ref_ID_30}\newline
STs harvest energy from PTs
& Depends on the PPP of PTs
& Markov chain\newline
Homogeneous PPP
& Single slope path-loss model \newline
EH from CCI of STs not mentioned
& Challenge \ref{chal1} and \ref{chal2}\newline
Maximize spatial throughput of SN
\\
\hline
Point-to-point uplink in wireless sensor networks (WSNs) \cite{Ref_ID_9}
&  Depends on spatial distribution of cellular transmitters (e.g., DPP)
& Ginibre point process
& Fixed transmit power for ambient RF source
& Challenge \ref{chal1} and \ref{chal2}\newline
  Derive the upper bound of power and transmission outage probabilities.
\\
\hline
Simultaneous wireless information
and power transfer (SWIPT) in WSNs for downlink \cite{Ref_ID_28}
& Depends on spatial distribution of transmitters (DPP)
& Ginibre point process
& Fixed transmit power for ambient RF source\newline
& Challenge \ref{chal1}\newline
Derive upper bound for power and transmission outage probability
\\
\hline
EH from CCI in a point-to-point wireless link \cite{Ref_ID_42}
& Arbitrary random variable\newline(Exponential distribution)
&  Optimization (heuristic)
& Arbitrary model of energy arrival\newline
Limited to single-user setup
&Challenge \ref{chal4} \newline 
Derive the optimal switching mode
\\
\hline
Ambient RF EH in a point-to-point sensor network~\cite{OPOW2015}
& Stochastic \newline
& Order Statistics  
& Single-slope path-loss model \newline
Fixed distance between sensor and sink and interfering BS.
& Challenge \ref{chal3} \newline 
Derive closed-form expression for the distribution
function of harvested energy.
\\
\hline
Ambient EH in a point-to-point wireless link \cite{EHWC2015}
& Deterministic \& stochastic \newline(Gamma distribution)
&  Stochastic optimization
& Co-tier interference and \newline
effect of path-loss are neglected
& Challenge \ref{chal4} \newline 
Active ratios of the transmitter and the receiver; 
derive energy cooperation strategies.
\\
\hline
\end{tabular}
\label{compare:association}
\end{center}
\end{table*}
\end{savenotes}

One of the main limitations of the existing performance analysis/feasibility studies is the use of single-slope path-loss models as it might not accurately capture the impact of near and far interfering distances. 
Moreover, the  energy arrival models need to capture  the impact of intensity, locations, and coordination of the interferers. Also, the rate-energy trade-off analysis is usually conducted without considering the effect of CCI which would be a significant issue in ultra-dense SCNs. 

\section*{Feasibility of RF-Based Ambient Energy Harvesting SCNs}
\label{forth:feasibility}
Although CCI is one of the main limitations of ultra-dense SCNs, it can be beneficial for energy harvesting. Therefore, dense deployment of SBSs and RF-based ambient energy harvesting are perfect complement to each other. In this section, we quantitatively analyze the feasibility of RF-based ambient energy harvesting SCNs and observe the relevant trade-offs in terms of the SINR outage probability and the energy efficiency of downlink transmission to a typical user.
Note that SINR outage probability is an important metric from the  perspective of users, whereas energy efficiency is important from the perspective of both the users and network operators  since it considers both the transmission rate of a user and grid power consumption of the corresponding serving SBS.

\subsection*{Network Model and Performance Metrics}

The network model under consideration is composed of macro BSs along with a mixture of on-grid and off-grid SBSs. The proportion of on-grid and off-grid SBSs is denoted as $\beta$. Macro BSs are powered only by grid power while SBSs can operate in either on-grid or off-grid mode. We assume that the macro BSs and SBSs are spatially distributed according to two independent homogeneous PPPs $\Phi_{m}$  and $\Phi_{s}$ with spatial intensities  $\lambda_{m}$ and $\lambda_{s}$, respectively. By independent thinning of the PPP of SBSs, the on-grid and off-grid SBSs follows two other PPPs denoted by $ \Phi_a$  and  $\Phi_d$ with intensities $\beta \lambda_{s}$ and $(1-\beta) \lambda_{s}$, respectively. 

Since the desired link and interfering link distances are relatively short in the dense deployment of SCNs, conventional single-slope path-loss models (that cannot distinguish between near and far distances) may not be proper for system performance analysis \cite{Ref_ID_35}. As such, in this article, we consider a
dual-slope path-loss model and investigate its accuracy in modeling the performance of ultra-dense SCNs. If we denote the Euclidean distance between the transmitter and receiver by~$d_{t,r}$, we can define the dual-slope path-loss $L_{t,r}$ as follows~\cite{Ref_ID_35}: 
\[ L_{t,r} = \begin{cases} d_{t,r} ^{-\alpha_1}, & d_{t,r} > d_c \\ d_{t,r} ^{-\alpha_2}, &   d_{t,r} \leq d_c \end{cases} \]
where $d_c$ is the critical distance, $\alpha_1$ and $\alpha_2$ are the path-loss exponents for the near and far distances, respectively.
The performance metrics can then be described as follows:

\subsubsection*{SINR outage probability, $P_{\mathrm{out}}$} 
The probability that the downlink SINR of a typical user $(\mathrm{SINR}_u)$ is less than its desired target SINR value $(\theta_t)$ is referred to as SINR outage probability. Mathematically, it can be defined as
\begin{equation}
P_{\mathrm{out}} = \mathrm{Pr} \left(\mathrm{SINR}_u \leq \theta_{t} \right) = \mathrm{Pr}\left( \frac{p_s L_{u,s}}{I_u + N_0} \leq\theta_{t} \right)
\end{equation}
where $N_0$, $p_s$, $L_{u,s}$, $I_u$ are the noise power, transmit power of the serving SBS, path-loss between the typical user and its serving SBS, and aggregate interference at the user end, respectively. The user can be served solely by its closest SBS $s$. Note that in the case of on-grid SBS $p_{s} = P_s$. 

For off-grid SBSs, the transmit power depends on the amount of power harvested from the macro BSs and on-grid SBSs\footnote{We consider instantaneous energy harvesting that has bursty nature and the effect of recharging time is neglected.}. This power cannot exceed the battery capacity $ P_s$, i.e.,
\begin{equation}
p_s  = \min \Bigg(P_s,\eta \Big(\sum_{n \in \Phi_a} p_{n} L_{s,n} + \sum_{m \in \Phi_m 
 }p_{m} L_{s,m} \Big) \Bigg)
\end{equation}
where $\eta$ is RF-to-DC conversion efficiency, $p_{n}$ is transmit power of $n^{\mathrm{th}}$ on-grid SBS,  $p_m$ is the transmit power of $m^{\mathrm{th}}$ macro BS,  $L_{s,n}$ and $L_{s,m}$ represent the path-loss between SBS $s$ and $n^{\mathrm{th}}$ SBS and $m^{\mathrm{th}}$ macro BS, respectively.  The aggregate interference at the user can then be given as follows:
\begin{equation}
I_u = \sum_{n \in \Phi_s \setminus \{s\}}  p_n L_{u,n} + \sum_{m \in \Phi_m} p_m L_{u,m}.
\end{equation}

\subsubsection*{Energy efficiency (EE)} It is defined as the ratio of achievable data rate of a typical user and its corresponding grid power consumption. This is calculated as follows:  
\begin{equation}
\mathrm{EE} = \begin{cases}
\frac{\text{log}_2(1+\mathrm{SINR}_u)}{P_s + P_{\epsilon}}, & \text{$s \in \Phi_{a}$}\\ \\
\frac{\text{log}_2(1+\mathrm{SINR}_u)}{P_{\epsilon}}, &\text{$s \in \Phi_d$}
\end{cases}
\end{equation}
where $\text{log}_2(1+\mathrm{SINR}_u)$ is the data rate of the user and $P_{\epsilon}$ is a fixed static power consumption at a given serving SBS.

\subsection*{Numerical Results}

In our simulation setup, the transmit power for the  macro BSs and SBSs are taken as $P_m = 40$~dBm and $P_s = 23$~dBm, respectively \cite{Ref_ID_25}. The RF-DC conversion efficiency is reported to vary in between 50\%-80\% \cite{Ref_ID_33, Ref_ID_18}; therefore, we consider $\eta = 0.7$. The simulation values for noise spectral density, desired SINR threshold, and static power consumption are set to $N_0 = -120$ dBm, $\theta_t = 5$ dB, and $P_{\epsilon} = 6$~dBm, respectively.  

\subsubsection*{Trade-off related to increasing the intensity of SBSs in RF-based ambient energy harvesting SCNs}
\begin{figure}[hbpt]
  \begin{subfigure}{0.45\textwidth}
    \includegraphics[width=1.1\textwidth]{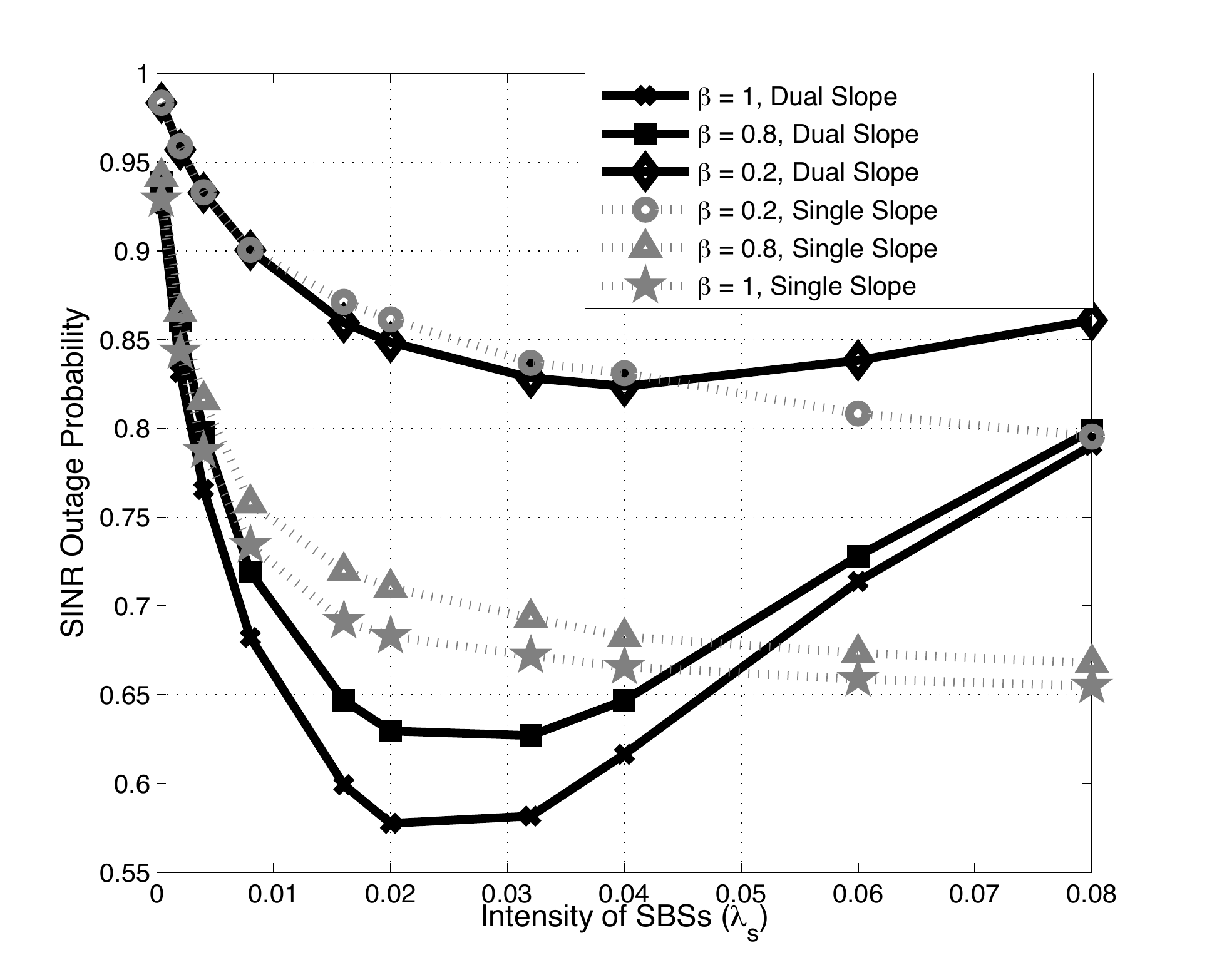}
    \vspace{-.15in}
    \caption{SINR outage probability of the typical user versus density of SBSs $\lambda_s $ ($\lambda_s = 50 \lambda_m$) with dual slope path-loss model where $\alpha_1$ = 4, $\alpha_2$ = 2, $d_c$ = 4 $m$ (single slope $\alpha_1= \alpha_2 = 4$).}
    \label{fig:3a}
  \end{subfigure}
  \begin{subfigure}{0.45\textwidth}
    \includegraphics[width=1.1\textwidth]{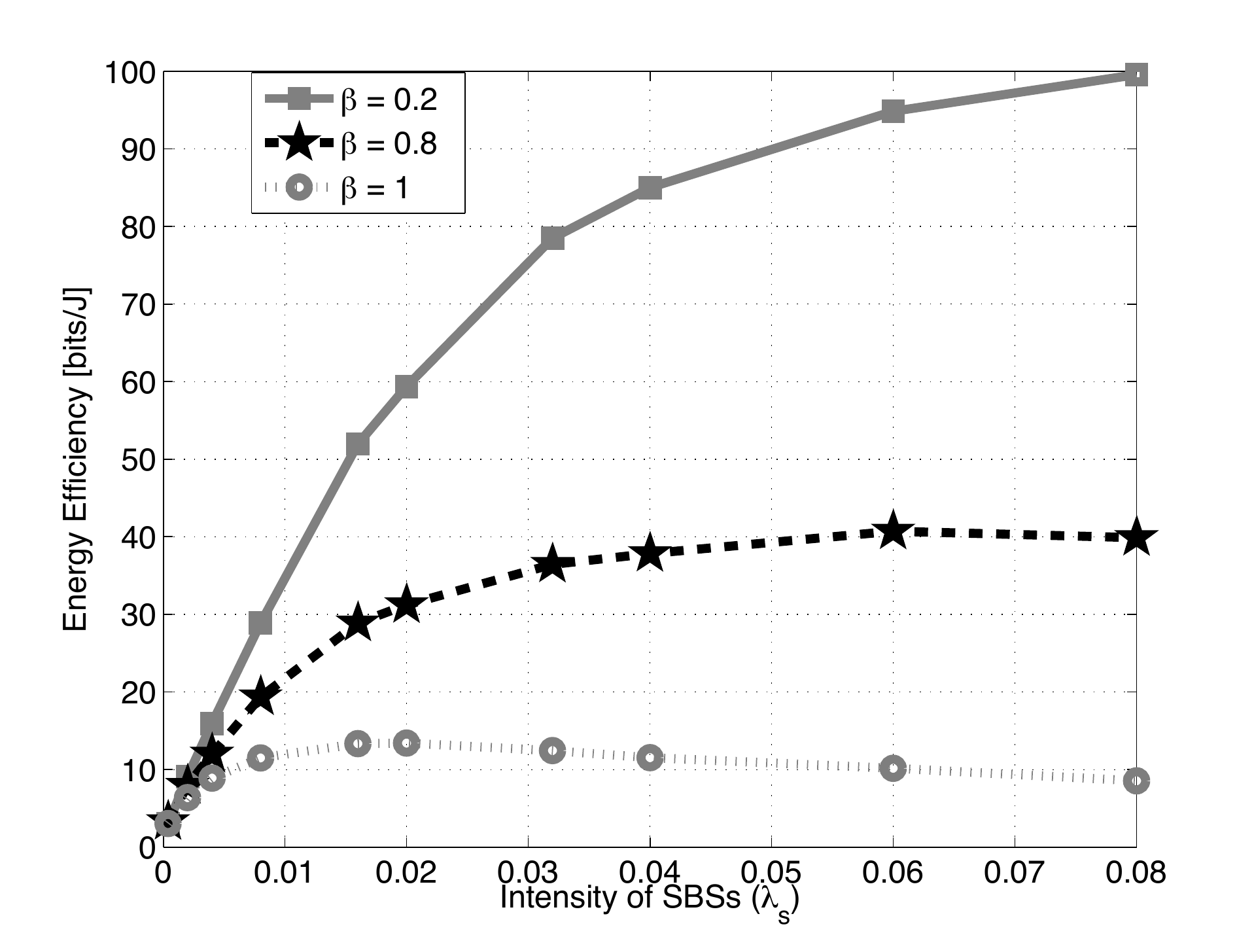}
    \vspace{-.15in}
    \caption{Energy efficiency of the typical user versus  $\lambda_s $ ($\lambda_s = 50 \lambda_m$) with dual slope path-loss model where $\alpha_1$ = 4, $\alpha_2$ = 2, $d_c$ = 4 $m$.}
    \label{fig:3b}
  \end{subfigure}
  \caption{Effect of intensity on SINR outage probability and energy efficiency of  RF-based energy harvesting SCNs.}
\end{figure}

High CCI degrades the received SINR of a typical user in the downlink but, at the same time, also increases the harvested energy at SBSs.
To investigate this phenomenon more closely, we study the impact of the increasing spatial intensity of SBSs on the downlink SINR outage probability of the typical user. 
\begin{itemize}
\item Fig. \ref{fig:3a} shows that increasing the density of SBSs ($\lambda_s$) first reduces the SINR outage probability due to increased amount of harvested energy. However, once the harvested energy starts exceeding battery storage limit of SBS,  further increase in $\lambda_s$ results in high CCI. The SINR outage probability of the typical user thus starts increasing again. The results therefore show the trade-off between CCI and amount of harvested energy  on the user performance when the intensity of SBSs keep increasing.
\item  Note that, the single-slope path-loss model is not capable of showing this trade-off in this parameter setting.
\item It is also observed that the optimal $\lambda_s$ reduces with the increasing proportions of on-grid SBSs ($\beta$). The reason is that a  higher number of on-grid SBSs significantly increases the amount of harvested energy at off-grid SBSs. Thus a much reduced SINR outage probability can be achieved with low values of $\lambda_s$  at the cost of higher grid power consumption.  
\end{itemize}
Selecting an optimal $\lambda_s$ is thus crucial in optimizing the performance of users in RF-based ambient energy harvesting ultra-dense SCNs. This can be done by selecting the idle mode for a required number of SBSs in the network in order to optimize the intensity of active SBSs in the network.

Fig. \ref{fig:3b}~illustrates that increasing $\lambda_s$ first increases the energy efficiency of the typical user rapidly because the transmission rate of the user is increasing due to increase in  the amount of energy harvested. However, with further increase of $\lambda_s$, the data transmission rate of the user starts to decrease due to interference. Consequently, the energy efficiency  gain decreases. It is important to note that the rate degradation does not lead to energy efficiency degradation for less fraction of on-grid BSs. The reason is that the benefit of reduced grid power consumption is greater than the drawback of rate degradation. However, the energy efficiency degradation can be seen from the result for $\beta = 1$ (100\% on-grid) where the  grid power consumption is high. It can thus be concluded that increasing  $\lambda_s$  in an unplanned manner is not necessarily beneficial from the energy efficiency perspective, especially for high fractions of on-grid SBSs. 

\subsubsection*{Trade-off related to increasing the proportion of on-grid SBSs in RF-based ambient energy harvesting SCNs}
For a given intensity of SBSs ($\lambda_s$), the harvested energy at off-grid SBSs increases along with the proportion of on-grid SBSs ($\beta$) in the network which in turn improves the rate of a typical user. However, at the same time, increase of $\beta$ degrades the energy efficiency due to a higher grid power consumption.

\begin{figure}[hbpt]
  \begin{subfigure}{0.45\textwidth}
    \includegraphics[width=1.1\textwidth]{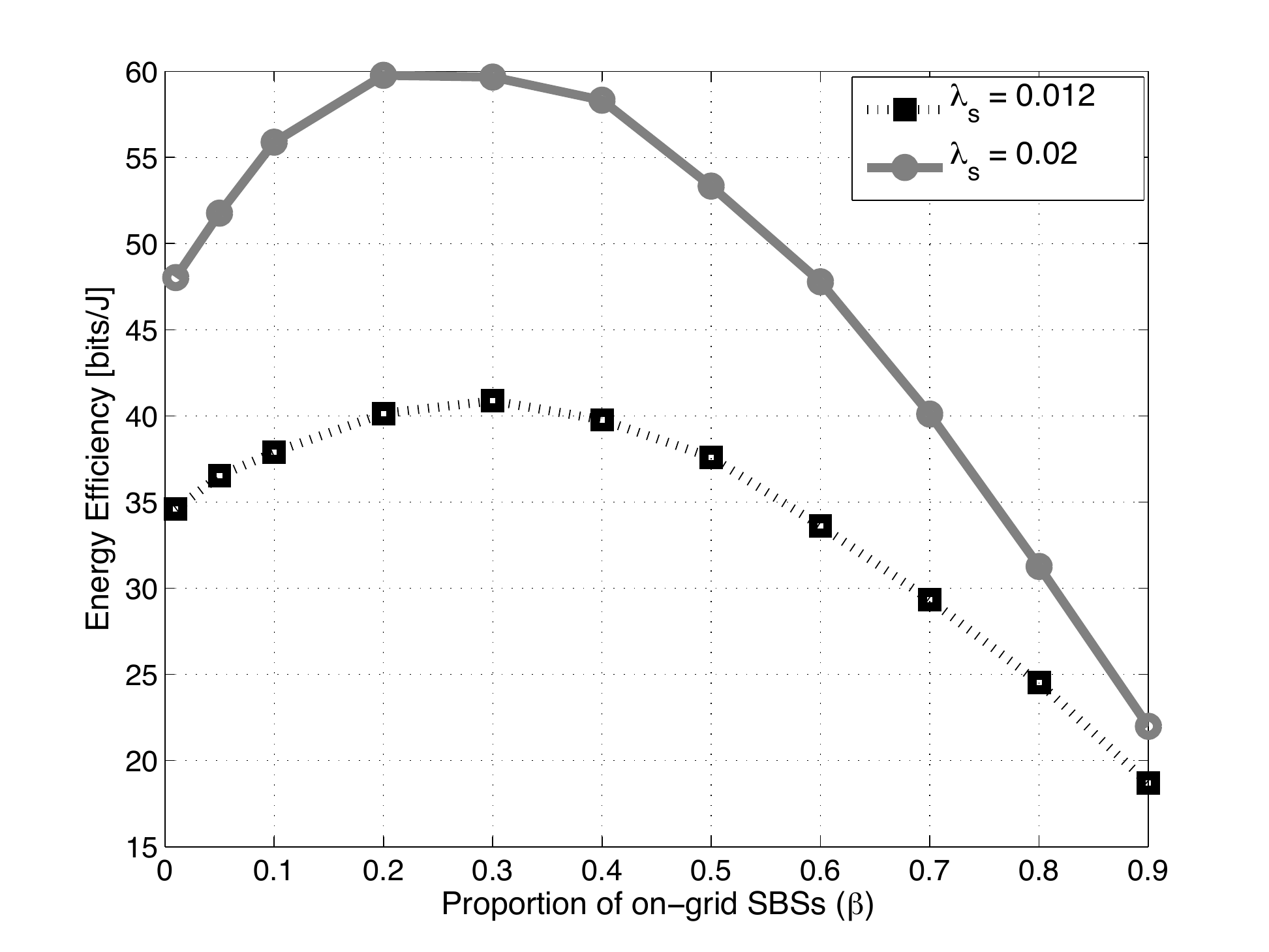}
    \caption{Energy efficiency of the {\em typical user} versus  $\beta$ with dual-slope path-loss model, where $\alpha_1$ = 4, $\alpha_2$ = 2, $d_c$ = 4 $m$.}
           \label{fig:4a}
  \end{subfigure}
  
  \begin{subfigure}{0.45\textwidth}
    \includegraphics[width=1.1\textwidth]{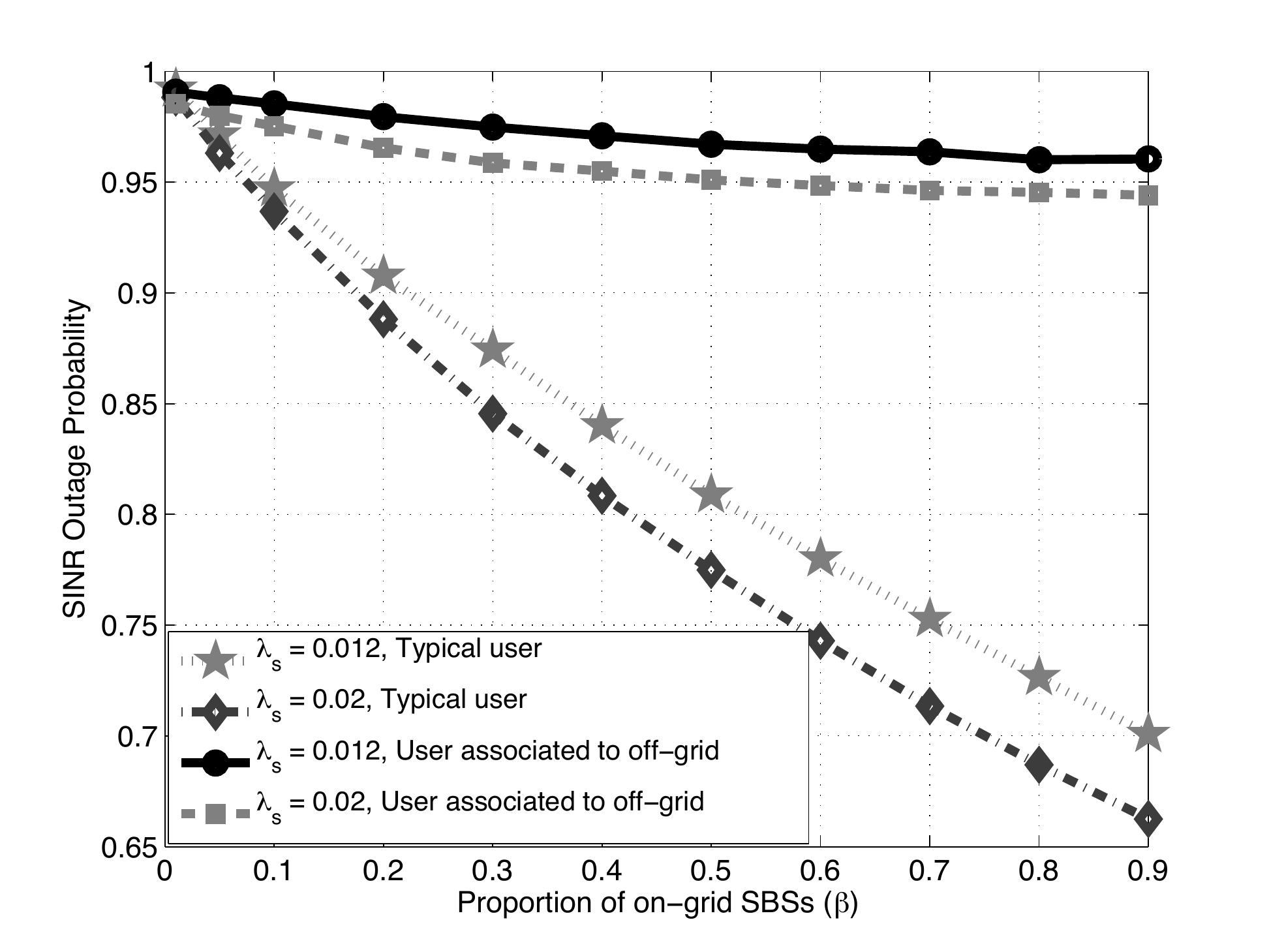}
    \caption{SINR outage probability of the {\em typical user} versus $\beta$ with dual-slope path-loss model, where $\alpha_1$ = 4, $\alpha_2$ = 2, $d_c$ = 4 $m$.}
    \label{fig:4b}
  \end{subfigure}
\caption{Effect of proportion of on-grid SBSs on SINR outage probability and energy efficiency of RF-based energy harvesting SCNs.}
\end{figure}

\begin{itemize}
\item Fig.~\ref{fig:4a} illustrates that increasing $\beta$, at first enhances the energy efficiency of a typical user due to increase in the amount of harvested energy and in turn the data transmission rate. But as we keep increasing $\beta$, grid power consumption becomes more dominant which consequently degrades energy efficiency. As such, the result shows the trade-off between the amount of harvested energy and grid power consumption as a function of $\beta$.
\item It can be observed that, for a given intensity of SBSs, there exists an optimal value of $\beta$ where the energy efficiency is maximum.
\item The optimal point, however, shifts, as the amount of harvested energy depends on $\lambda_s$ as well as $\beta$, i.e., for lower value of $\lambda_s$, we need more on-grid SBSs to reach the maximum energy efficiency.   Therefore, $\beta$ is an important design parameter to enhance the performance of RF-based ambient energy harvesting ultra-dense SCNs.
\end{itemize}

To show the SINR outage probability due to an off-grid SBS,  we consider the performance of a user who is always connected to the off-grid SBS. Then, we compare the performance results of such a user with those of a {\em typical user} who can be served by either an on-grid or an off-grid SBS. The simulation results for both cases are provided in Fig \ref{fig:4b}. We can see that for a given intensity of SBSs, increasing the proportion of on-grid SBSs decreases the SINR outage probability for both types of users.  
However, the performance gain of the {\em typical user} is significantly higher in comparison with the case that user is always connected to the off-grid SBSs. This implies that, for small cell users, flexible association should be better than off-grid only association.

\subsection*{Design Considerations}
\subsubsection*{Network-assisted user association}\label{chal5}
Due to varying energy levels and operating modes of different SBSs, user association in EH-enabled SCNs becomes more challenging compared to conventional grid-powered SCNs. A fully distributed user association criterion may not therefore be efficient for such networks since the users must consider the energy level, operating mode, and their required rate while making decision.  
Thus the user association criterion must be network-assisted in which the users can decide their associations by utilizing partial network information, e.g., traffic load, energy status, and the channel conditions of different SBSs.

\subsubsection*{Hybrid energy harvesting techniques} 
The amount of harvested energy level at some specific SBSs may not always be enough to handle their traffic load and it may also not be possible to offload their traffic to nearby SBSs. One potential solution to this problem is to exploit dedicated EH techniques.

\subsubsection*{Optimal use of harvested energy}
One future direction for the current work is to enable the storage for harvested energy and to find the best time and strategy for utilizing it. For example, SBSs with low traffic load can likely harvest and save ambient energy for  future use. Moreover, since the amount of harvested energy can be limited optimal/near-optimal power allocation policies for  SBSs are of prime importance. For instance, the optimal power of a given SBS should be selected such that both the amount of harvested energy at neighboring SBSs and the transmission requirements of the user associated to a given SBS can be fulfilled.

\subsubsection*{New analytical modeling/optimization tools}
In a generic setup for future networks, different types of SBSs based on the varying EH technologies such as ambient RF, dedicated RF, hybrid ambient RF-grid, hybrid dedicated RF-grid, hybrid ambient-dedicated RF, hybrid solar-grid, and hybrid solar-wind-grid will co-exist. These kind of heterogeneous EH networks result in different kind of user/ SBS clustering scenarios. To analyze these generic networks, it is crucial to use accurate propagation models and novel mathematical tools such as Ginibre point process, soft-core, and hard-core point process that can accurately capture repulsion (or attraction) between nodes and hot-spot scenarios. Also, to deal with the uncertainty of energy arrival rate, more sophisticated robust optimization techniques will be of significant importance.

\subsubsection*{Energy harvesting millimeter-wave (mm-wave) Networks}
Due to high bandwidth availability in mmWave bands, mmWave-based communication will be used in 5G networks \cite{Ref_ID_50}. An interesting future direction is to investigate EH-enabled mm-wave networks. The propagation model for these networks is quite different from those used in traditional cellular radio networks and operation of these networks requires highly directive transmissions. Performance modeling, analysis, and optimization of mmWave-based energy harvesting networks  is an interesting direction for future research.

\subsubsection*{Effect of path-loss and channel state information} 
The received power sensitivity of energy harvesting receiver is quite high compared to that of an information receiver.  Therefore, RF energy harvesting networks  are more sensitive to the effects of large-scale and small-scale fading, i.e., shadowing, multi-path fading, and distance attenuations.
Similar to information transmission network,  knowledge of channel state information (CSI)  plays an important role in RF-based EH networks, e.g., for cooperation among energy harvesting SBSs to share their energy states/requirements/transmission policies with each other and cognitive decision-making at SBSs about saving or utilizing their harvested energy.    

\section*{Conclusion}
\label{conclusion}
We have provided an overview of fundamental challenges of RF-based ambient energy harvesting ultra-dense SCNs. We have highlighted the existing techniques in the context of ambient energy harvesting  and their drawbacks. Then, we have presented a feasibility analysis of RF-based ambient energy harvesting SCNs. To show the potential trade-offs, we have investigated the effect of density of SBSs and proportion of on-grid SBS on SINR outage probability and energy efficiency of a typical user in the downlink.  Subsequently, we have shown that energy harvesting from co-channel interference is beneficial and the system parameters such as intensity of SBSs and proportion of on-grid SBSs can be optimized. 
\bibliographystyle{IEEEtran}

\begin{biography} 
{Amin Ghazanfari} received his B.Sc in Electrical Engineering from Islamic Azad University of Iran, in 2006. In 2014, he graduated from M.Sc in Wireless Communication Engineering from University of Oulu, Finland, with distinction. 
Currently, he is a Ph.D. Student in the department of Electrical and Computer Engineering, University of Manitoba, Canada. 
During the period of 2008 to 2011, he was working at the National
Telecommunication Institute, Iran. For his academic excellence, Amin has received several academic awards including trainee grant award and M.Sc thesis grant from Centre for Wireless Communications (CWC), Finland, 2012 and 2013, respectively.
\end{biography}

\begin{biography}
{Hina Tabassum} (M'13)
received the B.E. degree in electronic engineering from the NED University of
Engineering and Technology (NEDUET), Karachi, Pakistan, in 2004. She received during her undergraduate studies 2 gold medals from NEDUET
and SIEMENS for securing the first position among all engineering universities of Karachi. She then worked as lecturer in NEDUET for two
years. In September 2005, she joined the Pakistan Space and Upper Atmosphere Research Commission (SUPARCO), Karachi, Pakistan and
received there the best performance award in 2009. She completed her 
Masters and Ph.D. degree in Communications Engineering from NEDUET
in 2009 and King Abdullah University of Science and Technology (KAUST), 
Makkah Province, Saudi Arabia, in May 2013, respectively. Currently, she 
is working as a post-doctoral fellow in the University of Manitoba (UoM), 
Canada. Her research interests include wireless communications with focus on interference modeling, spectrum allocation, and power control in
heterogeneous networks.
\end{biography}

\begin{biography} 
{Ekram Hossain} (F?15) is a Professor in the Department of Electrical and Computer
Engineering at University of Manitoba, Winnipeg, Canada. He received his Ph.D. in Electrical
Engineering from University of Victoria, Canada, in 2001. Dr. Hossain's current research
interests include design, analysis, and optimization of wireless/mobile communications networks, cognitive
and green radio systems, and network economics. He has authored/edited several books in these areas
(http://home.cc.umanitoba.ca/$\sim$hossaina). He was elevated to an IEEE Fellow ``for contributions to spectrum management and resource allocation in cognitive and cellular radio networks".  Currently he serves as the Editor-in-Chief for the \textit{IEEE Communications Surveys and Tutorials} and an Editor for \textit{IEEE Wireless Communications}. Also, he is a member of the IEEE Press Editorial Board. Previously, he served as the Area Editor for the \textit{IEEE Transactions on Wireless Communications} in the area of ``Resource Management and Multiple Access'' from 2009-2011, an Editor for
the \textit{IEEE Transactions on Mobile Computing} from 2007-2012, and an Editor for the \textit{IEEE Journal on Selected Areas in Communications - Cognitive Radio Series} from 2011-2014. 
Dr. Hossain has won several research awards including
the IEEE Communications Society Transmission, Access, and Optical Systems (TAOS) Technical Committee's Best Paper Award in IEEE Globecom 2015, University of Manitoba Merit Award in 2010 and 2014 (for Research and
Scholarly Activities), the 2011 IEEE Communications Society Fred Ellersick
Prize Paper Award, and the IEEE Wireless Communications and Networking
Conference 2012 (WCNC'12) Best Paper Award. He is a Distinguished Lecturer of the
IEEE Communications Society (2012-2015). He is a registered Professional
Engineer in the province of Manitoba, Canada.
\end{biography}

\end{document}